\documentclass[twocolumn,showpacs,superscriptaddress,amsmath,amssymb]{revtex4-1}
\usepackage{graphicx}
\usepackage{CJKutf8}
\usepackage{dcolumn}
\usepackage{amsmath,bm}
\usepackage{epstopdf}
\usepackage[mathlines]{lineno}
\usepackage{color}
\usepackage[colorlinks=true,linkcolor=blue,citecolor=magenta,urlcolor=cyan]{hyperref}

\begin{document}
\title{The impacts of the quantum-dot confining potential on the spin-orbit effect}
\author{Rui\! Li}
\email{ruili@ysu.edu.cn}
\affiliation{Key Laboratory for Microstructural Material Physics of Hebei Province, School of Science, Yanshan University, Qinhuangdao 066004, China}
\affiliation{Quantum Physics and Quantum Information Division, Beijing Computational Science Research Center, Beijing 100193, China}

\author{Zhi-Hai\! Liu}
\affiliation{Quantum Physics and Quantum Information Division, Beijing Computational Science Research Center, Beijing 100193, China}

\author{Yidong\! Wu}
\affiliation{Key Laboratory for Microstructural Material Physics of Hebei Province, School of Science, Yanshan University, Qinhuangdao 066004, China}

\author{C.\!\! S.\! Liu}
\affiliation{Key Laboratory for Microstructural Material Physics of Hebei Province, School of Science, Yanshan University, Qinhuangdao 066004, China}

\begin{abstract}
For a nanowire quantum dot with the confining potential modeled by both the infinite and the finite square wells, we obtain exactly the energy spectrum and the wave functions in the strong spin-orbit coupling regime. We find that regardless of how small the well height is, there are at least two bound states in the finite square well: one has the $\sigma^{x}\mathcal{P}=-1$ symmetry and the other has the $\sigma^{x}\mathcal{P}=1$ symmetry. When the well height is slowly tuned from large to small, the position of the maximal probability density of the first excited state moves from the center to $x\ne0$, while the position of the maximal probability density of the ground state is always at the center.  A strong enhancement of the spin-orbit effect is demonstrated by tuning the well height. In particular, there exists a critical height $V^{c}_{0}$, at which the spin-orbit effect is enhanced to maximal.
\end{abstract}
\date{\today}
%\pacs{73.21.La, 71.70.Ej, 76.30.-v}
\maketitle

\section{Introduction}
The spin-orbit coupling (SOC), originating from the lacking of space-inversion symmetry in semiconductor materials~\cite{Winkler}, has played an important role in the studies of topological insulators~\cite{Hasan,Konig}, topological superconductors~\cite{Fu,Lutchyn,Oreg}, cold atom physics~\cite{YunLi,Chen,Mardonov}, spin quantum computings~\cite{Golovach,Tokura,Nadj,Nowak,Li,Hung}, etc. In the presence of SOC, the orbital degree of freedom of the electron is no longer separable from its spin degree of freedom, such that it is usually difficult to clarify the strong SOC effect in quantum system. It is also of fundamental interest to explore the physical properties of the quantum system beyond the weak SOC regime.

A semiconductor quantum dot~\cite{Wiel}, where a conduction electron of the material is localized by the nearby static electric gates, can be considered as an artificial atom. Unlike natural atoms, the artificial atom is more flexible because many system parameters are externally manipulable. The electronic~\cite{Reimann}, magnetic~\cite{Hanson}, and optical~\cite{Urbaszek} properties of the semiconductor quantum dot have attracted extensive research interest.

For quantum dot confined in quasi-2D with strong SOC, many theoretical works have devoted to solving the single electron energy spectrum. If the confining potential is of the cylindrical type, with the help of the Bessel function, one can get the exact energy spectrum~\cite{Bulgakov,Tsitsishvili,Intronati2013}. If the confining potential is of the harmonic type~\cite{Kuan2004,Lee2006,Rashba2,Bermeister}, there is no exact solution. For quantum dot confined in quasi-1D with strong SOC~\cite{Nowak,Li,Trif}, the situation would be a little different. Note that quantum dot with quasi-1D confinement, e.g., nanowire quantum dot~\cite{Berg,Stenhlik}, can already be fabricated experimentally. If the confining potential is of the harmonic type, the 1D quantum dot model can be mapped to the quantum Rabi model~\cite{Braak}, the energy spectrum can be solved using iteration method~\cite{Braak,Xie}.

In this paper, we study the strong spin-orbit effect in a quasi-1D quantum dot with the confining potential modeled by both the infinite square well (ISW) and the finite square well (FSW). With respect to both the $Z_{2}$ symmetry of the model and the energy region, we obtain a serious of transcendental equations, their solutions give rise to the exact energy spectrum of the quantum dot. The probability density distribution of the eigenstate in the FSW can be very different from that in the ISW. Interestingly, when we slowly lower the well height of the FSW, the position of the maximal probability density of the first excited state changes from the center to $x\ne0$; while the position of the maximal probability density of the ground state is always at the center. Finally, we study the electric-dipole transition rate between the lowest Zeeman sublevels. A strong enhancement of the transition rate by lowering the well height is demonstrated. In particular, we find that there exists a critical well height $V^{c}_{0}$, at which the spin-orbit effect is enhanced to maximal.

\section{The model}
\begin{figure}
\centering
\includegraphics{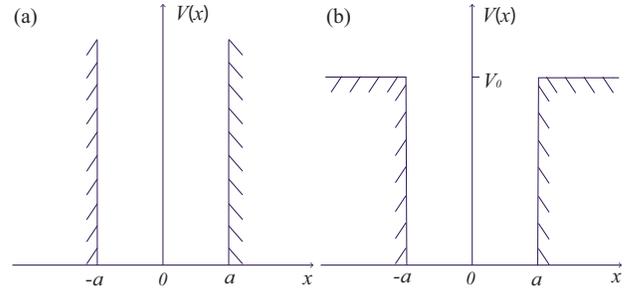}
\caption{\label{fig_model}Schematically shown the confining potential of a nanowire quantum dot. (a) ISW with width $a$. (b) FSW with width $a$ and height $V_{0}$.}
\end{figure}

We consider a model of nanowire quantum dot, where a conduction electron is confined in a 1D potential well and subject to both the Rashba spin-orbit field~\cite{bychkov} and the external Zeeman field. The Hamiltonian under consideration reads~\cite{Nowak,Echeverria, Gambetta} (we set $\hbar=1$)
\begin{equation}
H=\frac{p^{2}}{2m}+\alpha\sigma^{z}p+\Delta\sigma^{x}+V(x),\label{eq_model}
\end{equation}
where $m$ is the effective electron mass, $\alpha$ is the SOC strength, $\Delta=g_{e}\mu_{B}B/2$ is half of the Zeeman splitting induced by an external magnetic field $B$, and $V(x)$ is the confining potential. In this paper, we only focus on the strong SOC regime ($m\alpha^{2}>\Delta$), and the quantum-dot confining potential is modeled by both the ISW [see Fig.~\ref{fig_model}(a)] and the FSW [see Fig.~\ref{fig_model}(b)], i.e.,
\begin{equation}
V_{\rm I}(x)=\left\{
\begin{array}{ccc}
0 ,  & |x|<a,  \\
\infty,  & |x|>a,
\end{array}
\right.~
V_{\rm F}(x)=\left\{
\begin{array}{ccc}
0 ,  & |x|<a,  \\
V_{0},  & |x|>a,
\end{array}
\right.
\end{equation}
where $a$ and $V_{0}$ are the width and the height of the well, respectively.

Similar to the quantum Rabi model~\cite{Braak}, our model is also invariant under the following $Z_{2}$ transformation:
$(\sigma^{x}\mathcal{P})H(\sigma^{x}\mathcal{P})=H$,
where $\mathcal{P}$ is the parity operator. It follows that $\sigma^{x}\mathcal{P}$ and $H$ have common eigenfunction $\Psi(x)$, i.e., the eigenstates of the quantum dot can be specified with respect to the $Z_{2}$ symmetry. The $\sigma^{x}\mathcal{P}=1$ symmetry gives 
\begin{equation}
\Psi_{1}(x)=\Psi_{2}(-x),
\end{equation} and the $\sigma^{x}\mathcal{P}=-1$ symmetry gives  
\begin{equation}
\Psi_{1}(x)=-\Psi_{2}(-x),
\end{equation}
where $\Psi_{1,2}(x)$ are the two components of the eigenfunction $\Psi(x)=[\Psi_{1}(x),\Psi_{2}(x)]^{\rm T}$.

All the allowed energies of a quantum system are actually determined by its boundary condition. For the ISW [see Fig.~\ref{fig_model}(a)], the boundary condition simply reads
\begin{equation}
\Psi(a)=0.\label{eq_boundary}
\end{equation}
For the FSW [see Fig.~\ref{fig_model}(b)], the boundary condition reads
\begin{equation}
\Psi(a+0)=\Psi(a-0),~~~\Psi'(a+0)=\Psi'(a-0),\label{eq_boundary2}
\end{equation}
where $\Psi'(x)$ is the first derivative of the eigenfunction.  Note that the first equation is given by the continuous condition of the wave function and the second equation is given by the integration $\underset{\varepsilon\rightarrow\,0}{\rm lim}\int^{a+\varepsilon}_{a-\varepsilon}dx(H-E)\Psi(x)=0$ in the vicinity of the site $x=a$.

It should be noted that we do not need to consider the boundary condition at the other site $x=-a$.  Because when the boundary condition [see Eq.~(\ref{eq_boundary}) or (\ref{eq_boundary2})] at one site $x=a$ is satisfied, the boundary condition at the other site $x=-a$ is naturally satisfied due to the $Z_{2}$ symmetry. It should be also noted that, in our following calculations, we have chosen  InSb as our nanowire material. Unless otherwise stated, the model parameters are given in Table.~\ref{tab}.

\begin{table}
\centering
\caption{\label{tab}The parameters of a 1D InSb quantum dot used in our calculations ($m_{0}$ is the electron mass).}
%\begin{ruledtabular}
\begin{tabular}{|c|c|c|c|c|c|}
\hline
$m_{e}/m_{0}$~\cite{Nadj}&$g$~\cite{Nowak}&$B$~(T)&$\alpha$~(eV \AA)&$a$~(nm)&$V_{0}$~(meV) \\
\hline
$0.0136$&$50.6$&$0.8$&1$\sim$4&$50$&1.38\\
\hline
\end{tabular}
%\end{ruledtabular}
\end{table}

\section{The bulk spectrum and the bulk wave functions}
Because of the specific form of the confining potential $V(x)$ [see Fig.~\ref{fig_model}], the Hamiltonian $H$ can be reduced to either $H_{\rm b}=\frac{p^{2}}{2m}+\alpha\sigma^{z}p+\Delta\sigma^{x}$ (inside the well) or $H_{\rm b}+V_{0}$ (outside the well). In order to find the energy spectrum and the corresponding wave functions of our model, we first study the properties of the bulk Hamiltonian $H_{\rm b}$.

The bulk spectrum and the corresponding bulk wave functions in the energy region $E_{\rm b}\ge-\frac{1}{2}m\alpha^{2}-\frac{\Delta^{2}}{2m\alpha^{2}}$ can be found elsewhere~\cite{Li3}. The bulk spectrum of plane-wave solution reads~\cite{Li3}
\begin{equation}
E^{\pm}_{\rm b}=\frac{k^{2}}{2m}\pm\sqrt{\alpha^{2}k^{2}+\Delta^{2}}.\label{eq_bulkspectrumI}
\end{equation}
%The corresponding bulk wave functions read
%\begin{equation}
%\Psi^{+}_{\rm b}=\left\{\begin{array}{c}e^{ikx}\left(\begin{array}{c}\cos\frac{\theta}{2}\\\sin\frac{\theta}{2}\end{array}\right)\\
%e^{-ikx}\left(\begin{array}{c}\sin\frac{\theta}{2}\\\cos\frac{\theta}{2}\end{array}\right)\end{array}\right.
%\Psi^{-}_{\rm b}=\left\{\begin{array}{c}e^{ikx}\left(\begin{array}{c}\sin\frac{\theta}{2}\\-\cos\frac{\theta}{2}\end{array}\right)\\
%e^{-ikx}\left(\begin{array}{c}\cos\frac{\theta}{2}\\-\sin\frac{\theta}{2}\end{array}\right)\end{array}\right.,\label{eq_branchwave1}
%\end{equation}
%where $\theta\equiv\theta(k)=\arctan\left[\Delta/(\alpha\,k)\right]$. 
The bulk spectrum of exponential-function solution reads~\cite{Li3}
\begin{equation}
E^{\pm}_{\rm b}=-\frac{\Gamma^{2}}{2m}\pm\sqrt{-\alpha^{2}\Gamma^{2}+\Delta^{2}}.\label{eq_bulkspectrumII}
\end{equation}
%The corresponding bulk wave functions read
%\begin{equation}
%\Psi^{+}_{\rm b}=\left\{\begin{array}{c}e^{-\Gamma\,x}\left(\begin{array}{c}e^{i\varphi}\\1\end{array}\right)\\
%e^{\Gamma\,x}\left(\begin{array}{c}e^{-i\varphi}\\1\end{array}\right)\end{array}\right.
%\Psi^{-}_{\rm b}=\left\{\begin{array}{c}e^{-\Gamma\,x}\left(\begin{array}{c}-e^{-i\varphi}\\1\end{array}\right)\\
%e^{\Gamma\,x}\left(\begin{array}{c}-e^{i\varphi}\\1\end{array}\right)\end{array}\right.,\label{eq_branchwave2}
%\end{equation}
%where $\varphi\equiv\varphi(\Gamma)=\arctan\left(\alpha\Gamma/\sqrt{-\alpha^{2}\Gamma^{2}+\Delta^{2}}\right)$.
Inside the well $|x|<a$, the eigenfunction $\Psi(x)$ of Hamiltonian (\ref{eq_model}) can be expanded in terms of the four degenerate bulk wave functions~\cite{Li3}. However, outside the well $|x|>a$ for the FSW (classical forbidden region), the electron must have a dissipative energy $E_{\rm b}<-\frac{1}{2}m\alpha^{2}-\frac{\Delta^{2}}{2m\alpha^{2}}$, otherwise, the bound state can not be formed.  In the following, we address the bulk spectrum and the corresponding bulk wave functions in the dissipative energy region. The bulk wave function in this region can be assumed as
\begin{equation}
\Psi_{\rm b}(x)=\left(
\begin{array}{c}
\chi_{1} \\
\chi_{2}
\end{array}
\right)e^{ik_{\rho}e^{i\phi}x},
\end{equation}
where $k_{\rho}e^{i\phi}$ is a general complex number with amplitude $k_{\rho}$ and phase $\phi$. This solution can also be considered as a combined plane-wave and exponential-function solution. Substituting the bulk wave function $\Psi_{\rm b}(x)$ in Schr\"odinger equation $(H_{\rm b}-E_{\rm b})\Psi_{\rm b}(x)=0$ with the above expression, we have
\begin{equation}
\left(
\begin{array}{cc}
\frac{k^{2}_{\rho}e^{2i\phi}}{2m}+\alpha\,k_{\rho}e^{i\phi}-E_{\rm b}  & \Delta \\
\Delta  &  \frac{k^{2}_{\rho}e^{2i\phi}}{2m}-\alpha\,k_{\rho}e^{i\phi}-E_{\rm b}
\end{array}
\right)\cdot\left(
\begin{array}{c}
\chi_{1} \\
\chi_{2}
\end{array}
\right)=0.\label{eq_bulkspectrum}
\end{equation}
Letting the determinant of the matrix (the left $2\times2$ matrix) equal to zero, we have the following two coupled equations
\begin{eqnarray}
k^{2}_{\rho}\cos2\phi&=&2m(E_{b}+m\alpha^{2}),\nonumber\\
k^{4}_{\rho}&=&4m^{2}(E^{2}_{\rm b}-\Delta^{2}).
\end{eqnarray}
Combining these two equations and eliminating the variable $k_{\rho}$, we obtain the bulk spectrum
\begin{equation}
\frac{E^{\pm}_{\rm b}}{m\alpha^{2}}=\frac{-1\pm\sqrt{(1-\frac{\Delta^{2}}{m^{2}\alpha^{4}}\sin^{2}2\phi)\cos^{2}2\phi}}{\sin^{2}2\phi}.\label{eq_bulkspectrumIII}
\end{equation}
Once the bulk energy $E_{\rm b}$ is obtained, we can obtain four degenerate bulk wave functions via Eq.~(\ref{eq_bulkspectrum})
\begin{eqnarray}
\Psi^{1,3}_{\rm b}(x)&=&\left(
\begin{array}{c}
 1   \\
  R\,e^{\pm\,i\Phi}
\end{array}
\right)e^{ik_{\rho}x\cos\phi\mp\,k_{\rho}x\sin\phi},\nonumber\\
\Psi^{2,4}_{\rm b}(x)&=&\left(
\begin{array}{c}
 R\,e^{\mp\,i\Phi}   \\
  1
\end{array}
\right)e^{-ik_{\rho}x\cos\phi\mp\,k_{\rho}x\sin\phi},
%\Psi^{3}_{\rm b}(x)&=&\left(
%\begin{array}{c}
% 1   \\
%  R\,e^{-i\Phi}
%\end{array}
%\right)e^{ik_{\rho}x\cos\phi+k_{\rho}x\sin\phi},\nonumber\\
%\Psi^{4}_{\rm b}(x)&=&\left(
%\begin{array}{c}
% R\,e^{i\Phi}   \\
 % 1
%\end{array}
%\right)e^{-ik_{\rho}x\cos\phi+k_{\rho}x\sin\phi},\label{eq_branchwave3}
\end{eqnarray}
where
\begin{eqnarray}
R\cos\Phi&=&-\frac{m\alpha^{2}+\alpha\,k_{\rho}\cos\phi}{\Delta},\nonumber\\
R\sin\Phi&=&-\frac{k^{2}_{\rho}\sin2\phi+2m\alpha\,k_{\rho}\sin\phi}{2m\Delta}.
\end{eqnarray}
Outside the well $|x|>a$ (classical forbidden region), the eigenfunction $\Psi(x)$ of Hamiltonian (\ref{eq_model}) can be expanded in terms of the above four degenerate bulk wave functions.

\begin{figure}
\centering
\includegraphics[width=8.5cm]{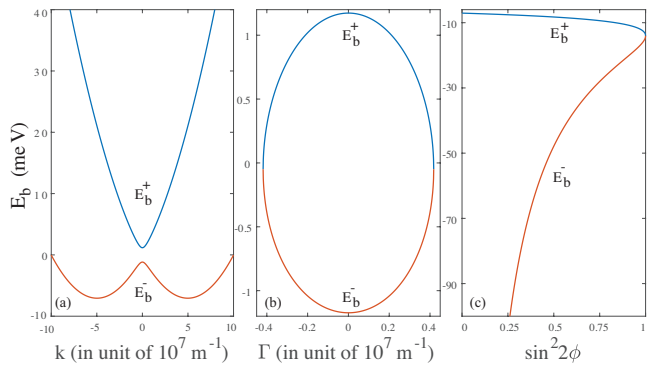}
\caption{\label{fig_bulkspectrum}The bulk spectrum of the quantum dot with strong SOC $\alpha=2.8$ eV \AA.  (a) The bulk spectrum of plane-wave solution (\ref{eq_bulkspectrumI}). (b) The bulk spectrum of exponential-function solution (\ref{eq_bulkspectrumII}). (c) The bulk spectrum of combined plane-wave and exponential-function solution (\ref{eq_bulkspectrumIII}).}
\end{figure}

Here, taking the InSb nanowire quantum dot as an example, we give the bulk spectrum of the Hamiltonian in the strong SOC regime ($m\alpha^{2}>\Delta$). Figures~\ref{fig_bulkspectrum}(a), (b), and (c) respectively show the bulk spectrum of the plane-wave, the exponential-function, and the combined plane-wave and exponential-function solutions. Also, from the detailed expressions of the bulk spectrum given in Eqs.~(\ref{eq_bulkspectrumI}), (\ref{eq_bulkspectrumII}), and (\ref{eq_bulkspectrumIII}), we have the following general results which are very useful for the following discussions. For the plane-wave solution [see Fig.~\ref{fig_bulkspectrum}(a)],  $E^{+}_{{\rm b}}\ge\Delta$ and $E^{-}_{\rm b}\ge-\frac{1}{2}m\alpha^{2}-\frac{\Delta^{2}}{2m\alpha^{2}}$. For the exponential-function solution [see Fig.~\ref{fig_bulkspectrum}(b)], $-\frac{\Delta^{2}}{2m\alpha^{2}}\le\,E^{+}_{\rm b}\le\Delta$ and $-\Delta\le\,E^{-}_{\rm b}\le\,-\frac{\Delta^{2}}{2m\alpha^{2}}$. For the combined solution [see Fig.~\ref{fig_bulkspectrum}(c)], $-m\alpha^{2}\le\,E^{+}_{\rm b}\le\,-\frac{1}{2}m\alpha^{2}-\frac{\Delta^{2}}{2m\alpha^{2}}$ and $E^{-}_{\rm b}\le\,-m\alpha^{2}$.

\section{The energy spectrum and the wave functions}
\begin{figure}
\centering
\includegraphics{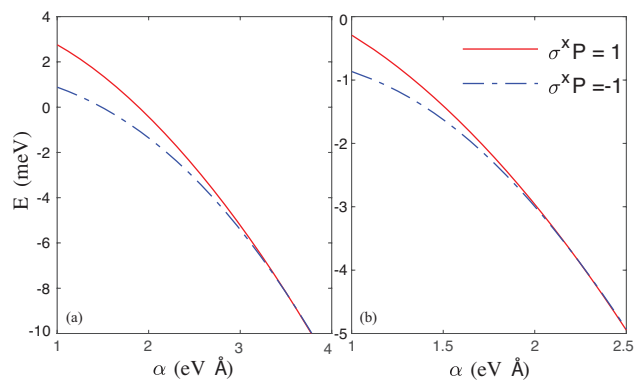}
\caption{\label{fig_infinitespectrum}The lowest two energy levels as a function of the SOC strength $\alpha$. (a) The results in the ISW. (b) The results in the FSW.}
\end{figure}

Since the bulk spectrum and the corresponding bulk wave functions of our model are obtained, the calculations for the energy spectrum are straightforward. The eigenfunction $\Psi(x)$ of Hamiltonian (\ref{eq_model}) is expanded in terms of the degenerate bulk wave functions~\cite{Bulgakov,Intronati2013,Kuan2004}. Imposing proper boundary condition [see Eq.~(\ref{eq_boundary}) or (\ref{eq_boundary2})] on $\Psi(x)$, we analytically derive a series of transcendental equations with respect to both the $Z_{2}$ symmetry and the energy region (for details see Appendix~\ref{appendix_b} and~\ref{appendix_c}). The solutions of these transcendental equations give us the exact energy spectrum.

Figures~\ref{fig_infinitespectrum}(a) and (b) show the two lowest energy levels as a function of the SOC $\alpha$ in the ISW and the FSW, respectively. First, with increasing the SOC, the effective Zeeman splitting becomes smaller, similar results were also obtained in a 2D quantum dot~\cite{Tsitsishvili}. In the large SOC limit $m\alpha^{2}\gg\Delta$, i.e., $\Delta\rightarrow0$, Hamiltonian (\ref{eq_model}) is time reversal invariant, hence each level is 2-fold degenerate due to Kramer's degeneracy. Second, the effective Zeeman splitting is much smaller (the spin-orbit effect is much stronger) in the FSW. The spin-orbit effect in the quantum dot can roughly be characterized by the relative parameter $\langle\,x\rangle/x_{\rm so}$~\cite{Trif,Li}, where $\langle\,x\rangle$ is the width of the wave function and $x_{\rm so}=\hbar/(m\alpha)$ is the spin-orbit length.  Obviously, $\langle\,x\rangle_{\rm F}$ is larger than $\langle\,x\rangle_{\rm I}$, hence the spin-orbit effect is much stronger in the FSW.

\begin{figure}
\centering
\includegraphics[width=8.5cm]{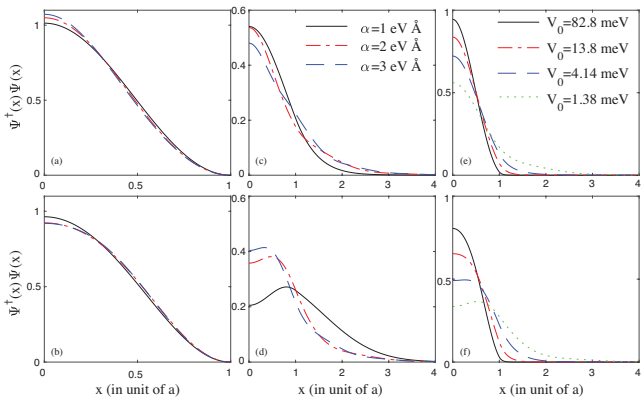}
\caption{\label{fig_wavefunction2}(a-d) The probability density distribution in both the ISW and the FSW with different SOC $\alpha$ (the height of the FSW is chosen as $V_{0}=1.38$ meV). (a) For the ground state in the ISW. (b) For the first excited state in the ISW. (c) For the ground state in the FSW. (d) For the first excited state in the FSW. (e-f) The probability density distribution in the FSW with different potential height $V_{0}$ (the SOC is chosen as $\alpha=1.8$ eV \AA). (e) For the ground state. (f) For the first excited state.}
\end{figure}

We also calculate the probability density distribution in the quantum dot for both the ground state and the first excited state. It should be noted that the Zeeman sublevels here are represented by the ground state and first excited state. Figures~\ref{fig_wavefunction2}(a) and (b) show the probability density distributions of the ground state and the first excited state in the ISW, respectively. Figures~\ref{fig_wavefunction2}(c) and~(d) show the probability density distributions of the ground state and the first excited state in the FSW, respectively. When the well height of the FSW is small $V_{0}=1.38$ meV, the probability density distribution in the FSW is apparently distinct from that in the ISW, where the position of the maximal probability density of the first excited state is not at the center [see Fig.~\ref{fig_wavefunction2}(d)].

It is of interest to know how $V_{0}$ affects the probability density distribution in the FSW. In Figs.~\ref{fig_wavefunction2}(e) and (f), for various well heights $V_{0}$, we show the probability density distributions of the ground and the first excited states respectively.  As can be seen from the figure, the position of the maximal probability density of the ground state is always at the center ($x=0$). When the well height is large, e.g., $V_{0}=82.8$ meV, the position of the maximal probability density of the first excited state is also at the center ($x=0$).  However, as we slowly lower $V_{0}$, there exists a critical $V^{c}_{0}$,  below which the position of the maximal probability density moves to $x\ne0$ [see Fig.~\ref{fig_wavefunction2}(f)]. This will induce interesting phenomena in the following discussion of the electric-dipole spin resonance.

We also find that no matter how small the well height $V_{0}$ is, there always exist at least two bound state in the FSW, one is labeled by the $\sigma^{x}\mathcal{P}=-1$ symmetry and the other is labeled by the $\sigma^{x}\mathcal{P}=1$ symmetry.

%\begin{figure}
%\centering
%\includegraphics{wavefunction.eps}
%\caption{\label{fig_wavefunction}The probability density distribution. (a) For the ground state in the ISW. (b) For the ground state in the FSW. (c) For the first excited state in the ISW. (d) For the first excited state in the FSW.}
%\end{figure}

\section{electric-dipole spin resonance}
In the presence of the SOC, the spin degree of freedom is mixed with the orbital degree of freedom, such that the spin in the quantum dot can respond to an external oscillating electric field $eEx\cos(\omega\,t)$, an effect called electric-dipole spin resonance~\cite{Golovach,Tokura,Nadj,Nowak,Li,Li4,Nowack,Laird,Rashba,Palyi,Romhanyi,ZhangPeng,Khomitsky,Ban}. Because the wave functions in the quantum dot are obtained in the previous section, we are able to calculate the Rabi frequency of the electric-dipole spin resonance.

When the frequency $\omega$ of the electric field matches the level spacing of the Zeeman sublevels, the electric field will induce an electric-dipole transition rate, i.e., the Rabi frequency, between the Zeeman sublevels
\begin{equation}
\Omega_{R}=2eE\Big|\int^{\Xi}_{0}dx\Psi^{\dagger}_{g}(x)x\Psi_{e}(x)\Big|,\label{eq_Rabi}
\end{equation}
where $\Xi=a$ and $\infty$ represent the integration boundary for the FSW and the ISW respectively, and $\Psi_{g,e}(x)$ denotes the ground (the first excited) state wave function. %Because of the $Z_{2}$ symmetry in our model, the integration can be simplified as
%\begin{equation}
%\Omega_{R}=2eE\Big|\int^{B}_{0}dx\Psi^{\dagger}_{g}(x)x\Psi_{e}(x)\Big|.
%\end{equation}
%It should noted that, the wave function should be normalized before calculation
%\begin{equation}
%\int^{B}_{-B}dx\Psi^{\dagger}(x)\Psi(x)=2\int^{B}_{0}dx\Psi^{\dagger}(x)\Psi(x)=1.
%\end{equation}
%All the calculations are done numerically.

\begin{figure}
\centering
\includegraphics[width=8.5cm]{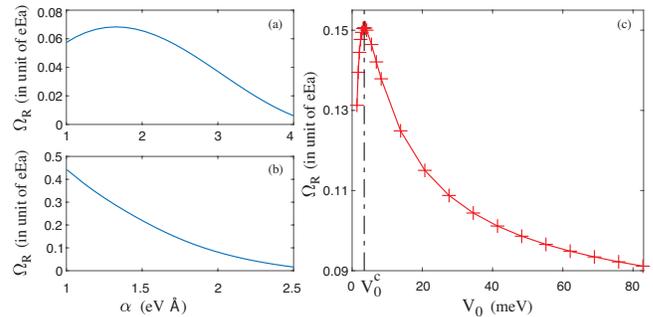}
\caption{\label{fig_transition}(a) The Rabi frequency as a function of the SOC in the ISW. (b) The Rabi frequency as a function of the SOC in the FSW. (c)The Rabi frequency as a function of the well height $V_{0}$ in the FSW. The SOC is fixed at $\alpha=1.8$ eV \AA .}
\end{figure}

In Figs.~\ref{fig_transition}(a) and (b), we show the Rabi frequency as a function of the SOC $\alpha$ in the ISW and the FSW, respectively. The Rabi frequency in the FSW can be almost one order larger than that in the ISW. Why the spin-orbit effect is so large in the FSW? We trace back to the wave functions given in Fig.~\ref{fig_wavefunction2}. In the FSW, the position of the maximal probability density of the first excited state is not at $x=0$, while the position of the maximal probability density of the ground state is at $x=0$, such that it is possible to produce a large Rabi frequency via Eq.~(\ref{eq_Rabi}). Also, in the large SOC limit $\alpha\rightarrow\infty$, the Rabi frequency becomes zero~\cite{Trif,Li}. This is because in the large SOC limit $m\alpha^{2}\gg\Delta$, i.e., $\Delta\rightarrow0$, the operator $\sigma^{z}$ would be a good quantum number [see Eq.~(\ref{eq_model})], hence the Rabi frequency is zero.

In Fig.~\ref{fig_transition}(c), we show the dependence of the Rabi frequency on the well height $V_{0}$ of the FSW. Obviously, in the large $V_{0}$ limit, e.g., $V_{0}\rightarrow\infty$, the Rabi frequency in the FSW would coincide with that in the ISW (see Figs.~\ref{fig_transition}). Lower the well height can remarkably enhance the SOC effect in the quantum dot. Interestingly, we find there exists a critical well height $V^{c}_{0}$, at which the Rabi frequency becomes maximal [see Fig.~\ref{fig_transition}(c)], i.e., the spin-orbit effect is enhanced to maximal. Below the critical $V^{c}_{0}$, if we continue to lower $V_{0}$, the Rabi frequency decreases sharply. This result is reasonable, it is impossible to infinitely enhance the spin-orbit effect. In the $V_{0}\rightarrow0$ limit, the Zeeman sublevels would become degenerate, such that there must exist a critical $V^{c}_{0}$ somewhere when we lower the well height.

\section{Summary}
In the presence of both the strong SOC and the Zeeman field, we have obtained exactly the energy spectrum and the corresponding wave functions in both the ISW and the FSW.  The spin-orbit effect is much stronger in the FSW than that in the ISW. Moreover, the probability density distribution in the FSW can be very different from that in the ISW. A strong enhancement of the SOC effect is demonstrated by tuning the height of the confining potential. In particular, we show that there exists a critical well height, at which the spin-orbit effect is enhanced to maximal.

\section*{Acknowledgements}
This work is supported by National Natural Science Foundation of China Grant No.~11404020 and Postdoctoral Science Foundation of China Grant No.~2014M560039.

\appendix
\section{\label{appendix_b}The transcendental equations in the ISW quantum dot}

%For a given bulk energy, there exists four degenerate bulk wave functions. Thus, the eigenfunction $\Psi(x)$ of the Hamiltonian can be expanded in terms of these degenerate bulk wave functions. Imposing proper boundary condition on the eigenfunction, we should have the transcendental equation which can determine the energy spectrum of our model.

\subsection{Energy region: $-\frac{1}{2}m\alpha^{2}-\frac{\Delta^{2}}{2m\alpha^{2}}\le\,E\le-\Delta$}
In this energy region, as can be seen from the bulk spectrum [see Fig.~\ref{fig_bulkspectrum}(a)], one can find four $k$ solutions $\pm\,k_{1,2}$ from the `$-$' dispersion relation given in Eq.~(\ref{eq_bulkspectrumI})
\begin{equation}
k_{1,2}=\sqrt{2}m\alpha\sqrt{1+\frac{E}{m\alpha^{2}}\pm\sqrt{1+2\frac{E}{m\alpha^{2}}+\frac{\Delta^{2}}{m^{2}\alpha^{4}}}}.
\end{equation}
Thus, the eigenfunction $\Psi(x)$ can been written as a linear combination of these four degenerate bulk wave functions. Note that all of the four bulk wave functions belong to the `$-$' branch.  In the coordinate region $|x|<a$, the eigenfunction reads~\cite{Li3}
\begin{eqnarray}
\Psi(x)&=&c_{1}\left(
\begin{array}{c}
\sin\frac{\theta_{1}}{2}   \\
-\cos\frac{\theta_{1}}{2}
\end{array}
\right)e^{ik_{1}x}+c_{2}\left(
\begin{array}{c}
\cos\frac{\theta_{1}}{2}   \\
-\sin\frac{\theta_{1}}{2}
\end{array}
\right)e^{-ik_{1}x}\nonumber\\
&&+c_{3}\left(
\begin{array}{c}
\sin\frac{\theta_{2}}{2}   \\
-\cos\frac{\theta_{2}}{2}
\end{array}
\right)e^{ik_{2}x}+c_{4}\left(
\begin{array}{c}
\cos\frac{\theta_{2}}{2}   \\
-\sin\frac{\theta_{2}}{2}
\end{array}
\right)e^{-ik_{2}x},\nonumber\\\label{eq_generalwfun1}
\end{eqnarray}
where $\theta_{1,2}=\arctan\big[\Delta/(\alpha\,k_{1,2})\big]$ and $c_{1,2,3,4}$ are the coefficients to be determined.

As we emphasized before, we can specify the eigenfunction $\Psi(x)$ with respect to the $Z_{2}$ symmetry. This symmetry gives some constraints on the coefficients. For the $\sigma^{x}\mathcal{P}=1$ symmetry, the relationship $\Psi_{1}(x)=\Psi_{2}(-x)$ gives rise to $c_{2}=-c_{1}$ and $c_{4}=-c_{3}$. For the $\sigma^{x}\mathcal{P}=-1$ symmetry, the relationship $\Psi_{1}(x)=-\Psi_{2}(-x)$ gives rise to $c_{2}=c_{1}$ and $c_{4}=c_{3}$. In other words, we only have two coefficients $c_{1,3}$ to be determined. Using the hard-wall boundary condition $\Psi_{1,2}(a)=0$, we obtain the following transcendental equation
\begin{equation}
\frac{\sin[(k_{1}-k_{2})a]}{\sin[(k_{1}+k_{2})a]}=\mp\,\frac{\sin[(\theta_{1}-\theta_{2})/2]}{\cos[(\theta_{1}+\theta_{2})/2]},\label{eq_tranI}
\end{equation}
where the minus sign `$-$' and the plus sign `$+$' correspond to the $\sigma^{x}\mathcal{P}=1$ and $\sigma^{x}\mathcal{P}=-1$ symmetries respectively.

\subsection{Energy region: $\Delta\le\,E$}
As also can be seen from the bulk spectrum [see Fig.~\ref{fig_bulkspectrum}(a)], one can find $\pm\,k_{2}$ and $\pm\,k_{1}$ solutions from the `$+$' and `$-$' dispersion relations given in Eq.~(\ref{eq_bulkspectrumI}), respectively.
%\begin{equation}
%k_{1}=\sqrt{2}m\alpha\sqrt{1+\frac{E}{m\alpha^{2}}-\sqrt{1+2\frac{E}{m\alpha^{2}}+\frac{\Delta^{2}}{m^{2}\alpha^{4}}}}.
%\end{equation}
%one also can find two solutions $\pm\,k_{1}$ from the `$-$' dispersion relation given in Eq.~(\ref{eq_bulkspectrum}).
%\begin{equation}
%k_{2}=\sqrt{2}m\alpha\sqrt{1+\frac{E}{m\alpha^{2}}+\sqrt{1+2\frac{E}{m\alpha^{2}}+\frac{\Delta^{2}}{m^{2}\alpha^{4}}}}.
%\end{equation}
The eigenfunction $\Psi(x)$ can be written as a linear combination of these four degenerate bulk wave functions, i.e., two from the `$+$' branch and two from the `$-$' branch. In the coordinate region $|x|<a$, the eigenfunction reads~\cite{Li3}
\begin{eqnarray}
\Psi(x)&=&c_{1}\left(
\begin{array}{c}
\cos\frac{\theta_{1}}{2}   \\
\sin\frac{\theta_{1}}{2}
\end{array}
\right)e^{ik_{1}x}+c_{2}\left(
\begin{array}{c}
\sin\frac{\theta_{1}}{2}   \\
\cos\frac{\theta_{1}}{2}
\end{array}
\right)e^{-ik_{1}x}\nonumber\\
&&+c_{3}\left(
\begin{array}{c}
\sin\frac{\theta_{2}}{2}   \\
-\cos\frac{\theta_{2}}{2}
\end{array}
\right)e^{ik_{2}x}+c_{4}\left(
\begin{array}{c}
\cos\frac{\theta_{2}}{2}   \\
-\sin\frac{\theta_{2}}{2}
\end{array}
\right)e^{-ik_{2}x}.\nonumber\\\label{eq_generalwfun2}
\end{eqnarray}
The $\sigma^{x}\mathcal{P}=1$ symmetry gives rise to $c_{2}=c_{1}$ and $c_{4}=-c_{3}$, and the $\sigma^{x}\mathcal{P}=-1$ symmetry gives rise to $c_{2}=-c_{1}$ and $c_{4}=c_{3}$, such that only two coefficients $c_{1,3}$ are to be determined. The hard-wall boundary condition $\Psi_{1,2}(a)=0$ gives us the following transcendental equation
\begin{equation}
\frac{\sin[(k_{1}+k_{2})a]}{\sin[(k_{1}-k_{2})a]}=\pm\frac{\sin[(\theta_{1}+\theta_{2})/2]}{\cos[(\theta_{1}-\theta_{2})/2]},\label{eq_tranII}
\end{equation}
where the plus sign `$+$' and the minus sign `$-$' correspond to the $\sigma^{x}\mathcal{P}=1$ and $\sigma^{x}\mathcal{P}=-1$ symmetries respectively.

\subsection{Energy region: $-\Delta\le\,E\le-\frac{\Delta^{2}}{2m\alpha^{2}}$}
In this energy region, one can find two solutions $\pm\,k$ from the `$-$' branch dispersion relation given in Eq.~(\ref{eq_bulkspectrumI})
\begin{equation}
k=\sqrt{2}m\alpha\sqrt{1+\frac{E}{m\alpha^{2}}+\sqrt{1+2\frac{E}{m\alpha^{2}}+\frac{\Delta^{2}}{m^{2}\alpha^{4}}}}.
\end{equation}
One also can find two solutions $\pm\Gamma$ from the `$-$' branch dispersion relation given in Eq.~(\ref{eq_bulkspectrumII})
\begin{equation}
\Gamma=\sqrt{2}m\alpha\sqrt{-1-\frac{E}{m\alpha^{2}}+\sqrt{1+2\frac{E}{m\alpha^{2}}+\frac{\Delta^{2}}{m^{2}\alpha^{4}}}}.
\end{equation}
Thus, in the coordinate region $|x|<a$, the eigenfunction $\Psi(x)$ can be expanded as a linear combination of the four degenerate bulk wave functions, i.e., two from the `$-$' branch of the plane-wave solution and two from the `$-$' branch of the exponential-function solution~\cite{Li3}
\begin{eqnarray}
\Psi(x)&=&c_{1}e^{-\Gamma\,x}\left(\begin{array}{c}-e^{-i\varphi}\\1\end{array}\right)+c_{2}e^{\Gamma\,x}\left(\begin{array}{c}-e^{i\varphi}\\1\end{array}\right)\nonumber\\
&&+c_{3}e^{ikx}\left(\begin{array}{c}\sin\frac{\theta}{2}\\-\cos\frac{\theta}{2}\end{array}\right)+c_{4}e^{-ikx}\left(\begin{array}{c}\cos\frac{\theta}{2}\\-\sin\frac{\theta}{2}\end{array}\right),\nonumber\\\label{eq_generalwfun3}
\end{eqnarray}
where $\theta\equiv\theta(k)=\arctan\left[\Delta/(\alpha\,k)\right]$ and $\varphi\equiv\varphi(\Gamma)=\arctan\left(\alpha\Gamma/\sqrt{-\alpha^{2}\Gamma^{2}+\Delta^{2}}\right)$.  For the $\sigma^{x}\mathcal{P}=1$ symmetry, the relationship $\Psi_{1}(x)=\Psi_{2}(-x)$ gives rise to $c_{2}=-c_{1}e^{-i\varphi}$ and $c_{4}=-c_{3}$. The boundary condition $\Psi_{1,2}(a)=0$ gives us the following transcendental equation
\begin{equation}
\frac{\cos(ka-\varphi/2)-e^{2\Gamma\,a}\cos(ka+\varphi/2)}{\cos(ka+\varphi/2)-e^{2\Gamma\,a}\cos(ka-\varphi/2)}=\tan(\theta/2).\label{eq_tranIII}
\end{equation}
For the $\sigma^{x}\mathcal{P}=-1$ symmetry, the relationship $\Psi_{1}(x)=-\Psi_{2}(-x)$ gives rise to $c_{2}=c_{1}e^{-i\varphi}$ and $c_{4}=c_{3}$. The boundary condition $\Psi_{1,2}(a)=0$ gives us the following transcendental equation
\begin{equation}
\frac{\sin(ka-\varphi/2)+e^{2\Gamma\,a}\sin(ka+\varphi/2)}{\sin(ka+\varphi/2)+e^{2\Gamma\,a}\sin(ka-\varphi/2)}=\tan(\theta/2).\label{eq_tranIV}
\end{equation}

\subsection{Energy region: $-\frac{\Delta^{2}}{2m\alpha^{2}}\le\,E\le\Delta$}
In this energy region, one can find two solutions $\pm\,k$ from the `$-$' branch dispersion relation given in Eq.~(\ref{eq_bulkspectrumI}).
One also can find two solutions $\pm\Gamma$ from the `$+$' branch dispersion relation of given in Eq.~(\ref{eq_bulkspectrumII}).
Thus, in the coordinate region $|x|<a$, the eigenfunction $\Psi(x)$ can be expanded as a linear combination of these four degenerate bulk wave functions, i.e., two from the `$-$' branch of the plane-wave solution and two from the `$+$' branch of the exponential-function solution~\cite{Li3}
\begin{eqnarray}
\Psi(x)&=&c_{1}e^{-\Gamma\,x}\left(\begin{array}{c}e^{i\varphi}\\1\end{array}\right)+c_{2}e^{\Gamma\,x}\left(\begin{array}{c}e^{-i\varphi}\\1\end{array}\right)\nonumber\\
&&+c_{3}e^{ikx}\left(\begin{array}{c}\sin\frac{\theta}{2}\\-\cos\frac{\theta}{2}\end{array}\right)+c_{4}e^{-ikx}\left(\begin{array}{c}\cos\frac{\theta}{2}\\-\sin\frac{\theta}{2}\end{array}\right).\nonumber\\\label{eq_generalwfun4}
\end{eqnarray}
For the $\sigma^{x}\mathcal{P}=1$ symmetry, the relationship $\Psi_{1}(x)=\Psi_{2}(-x)$ gives rise to $c_{2}=c_{1}e^{i\varphi}$ and $c_{4}=-c_{3}$. The boundary condition $\Psi_{1,2}(a)=0$ gives us the following transcendental equation
\begin{equation}
\frac{\sin(ka+\varphi/2)+e^{2\Gamma\,a}\sin(ka-\varphi/2)}{\sin(ka-\varphi/2)+e^{2\Gamma\,a}\sin(ka+\varphi/2)}=-\tan(\theta/2).\label{eq_tranV}
\end{equation}
For the $\sigma^{x}\mathcal{P}=-1$ symmetry, the relationship $\Psi_{1}(x)=-\Psi_{2}(-x)$ gives rise to $c_{2}=-c_{1}e^{i\varphi}$ and $c_{4}=c_{3}$. The boundary condition $\Psi_{1,2}(a)=0$ gives us the following transcendental equation
\begin{equation}
\frac{\cos(ka+\varphi/2)-e^{2\Gamma\,a}\cos(ka-\varphi/2)}{\cos(ka-\varphi/2)-e^{2\Gamma\,a}\cos(ka+\varphi/2)}=-\tan(\theta/2).\label{eq_tranVI}
\end{equation}

\section{\label{appendix_c}The transcendental equations in the FSW quantum dot}
Outside the square well $x>a$, because of the constraint ${\rm lim}_{x\rightarrow\infty}\Psi(x)=0$, the eigenfunction can only be written as
\begin{eqnarray}
\Psi(x)&=&c_{5}
\left(
\begin{array}{c}
 1   \\
R\,e^{i\Phi}
\end{array}
\right)e^{ik_{\rho}x\cos\phi-k_{\rho}x\sin\phi}\nonumber\\
&&+c_{6}\left(
\begin{array}{c}
 R\,e^{-i\Phi}   \\
1
\end{array}
\right)e^{-ik_{\rho}x\cos\phi-k_{\rho}x\sin\phi},\label{eq_generalwfun5}
\end{eqnarray}
where $c_{5,6}$ are the coefficients to be determined and
\begin{eqnarray}
k_{\rho}\cos\phi&=&m\alpha\sqrt{1+\frac{E-V_{0}}{m\alpha^{2}}+\sqrt{\frac{(E-V_{0})^{2}-\Delta^{2}}{m^{2}\alpha^{4}}}},\nonumber\\
k_{\rho}\sin\phi&=&m\alpha\sqrt{-1-\frac{E-V_{0}}{m\alpha^{2}}+\sqrt{\frac{(E-V_{0})^{2}-\Delta^{2}}{m^{2}\alpha^{4}}}}.\nonumber\\
\end{eqnarray}
The other two bulk wave functions $\Psi^{3,4}_{b}(x)$ are divergent in the limit $x\rightarrow\infty$. Inside the square well $|x|<a$, the eigenfunction can still be written as those given in Eqs.~(\ref{eq_generalwfun1}), (\ref{eq_generalwfun2}), (\ref{eq_generalwfun3}), and (\ref{eq_generalwfun4}) with respect to the energy region.

The eigenfunctions can still be specified with respect to the $Z_{2}$ symmetry.
%Note that in the coordinate region $x<-a$, the eigenfunction must be expanded in terms of $\Psi^{3,4}_{\rm b}(x)$ given in Eq.~(\ref{eq_branchwave3}), such that it can satisfy the constraint ${\rm lim}_{x\rightarrow-\infty}\Psi(x)=0$. Also, the expanded coefficients can be related to $c_{5,6}$ with respect to the $Z_{2}$ symmetry.
For eigenfunction inside the well, we have two coefficients $c_{1,3}$ to be determined. Also, for eigenfunction outside the well, we have the other two coefficients $c_{5,6}$ to be determined [see Eq.~(\ref{eq_generalwfun5})]. The boundary condition, given by Eq.~(\ref{eq_boundary2}), give us a matrix equation
\begin{equation}
\textbf{M}\cdot\textbf{C}=0.\label{eq_tranVII}
\end{equation}
where $\textbf{M}$ is a $4\times4$ matrix and $\textbf{C}=(c_{1},c_{3},c_{5},c_{6})^{\rm T}$. Let the determinant of the matrix $\textbf{M}$ equal to $0$, we obtain a transcendental equation which is an implicit equation of the energy $E$
\begin{equation}
{\rm det}(\textbf{M})=0.
\end{equation}  
Similar to the discussions in the ISW, here we also can obtain a series of transcendental equations with respect to both the $Z_{2}$ symmetry and the energy region.  The detailed expression of the matrix $\textbf{M}$ is given as followes. In the energy region $-\frac{1}{2}m\alpha^{2}-\frac{\Delta^{2}}{2m\alpha^{2}}\le\,E\le-\Delta$, the matrix $\textbf{M}$ reads
\begin{widetext}
\begin{equation}
M_{\pm}=\left(
\begin{array}{cccc}
 e^{ik_{1}a}\sin\frac{\theta_{1}}{2}\mp\,e^{-ik_{1}a}\cos\frac{\theta_{1}}{2} &  e^{ik_{2}a}\sin\frac{\theta2}{2}\mp\,e^{-ik_{2}a}\cos\frac{\theta2}{2} & -e^{ik_{x}a-k_{y}a}  & -R\times\\
 ~&~&~&e^{-i(k_{x}a+\Phi)-k_{y}a}\\
 -e^{ik_{1}a}\cos\frac{\theta_{1}}{2}\pm\,e^{-ik_{1}a}\sin\frac{\theta_{1}}{2} &  -e^{ik_{2}a}\cos\frac{\theta2}{2}\pm\,e^{-ik_{2}a}\sin\frac{\theta2}{2} & -R\times\, &-e^{-ik_{x}a-k_{y}a}\\
  ~&~&e^{i(k_{x}a+\Phi)-k_{y}a }&\\
 ik_{1}\left(e^{ik_{1}a}\sin\frac{\theta_{1}}{2}\pm\,e^{-ik_{1}a}\cos\frac{\theta_{1}}{2}\right)  & ik_{2}\left(e^{ik_{2}a}\sin\frac{\theta2}{2}\pm\,e^{-ik_{2}a}\cos\frac{\theta2}{2}\right)  &  -(ik_{x}-k_{y})\times &R(ik_{x}+k_{y})\times\\
  ~&~&e^{ik_{x}a-k_{y}a}&e^{-i(k_{x}a+\Phi)-k_{y}a}\\
-ik_{1}\left(e^{ik_{1}a}\cos\frac{\theta_{1}}{2}\pm\,e^{-ik_{1}a}\sin\frac{\theta_{1}}{2}\right) &-ik_{2}\left(e^{ik_{2}a}\cos\frac{\theta2}{2}\pm\,e^{-ik_{2}a}\sin\frac{\theta2}{2}\right)&-R(ik_{x}-k_{y})\times&(ik_{x}+k_{y})\times\\
 ~&~&e^{i(k_{x}a+\Phi)-k_{y}a }&e^{-ik_{x}a-k_{y}a}
\end{array}
\right).
\end{equation}
In the energy region $E\ge\Delta$, the matrix $\textbf{M}$ reads
\begin{equation}
M_{\pm}=\left(
\begin{array}{cccc}
 e^{ik_{1}a}\cos\frac{\theta_{1}}{2}\pm\,e^{-ik_{1}a}\sin\frac{\theta_{1}}{2} &  e^{ik_{2}a}\sin\frac{\theta2}{2}\mp\,e^{-ik_{2}a}\cos\frac{\theta2}{2} & -e^{ik_{x}a-k_{y}a}  & -R\times\\
 ~&~&~&e^{-i(k_{x}a+\Phi)-k_{y}a}\\
 e^{ik_{1}a}\sin\frac{\theta_{1}}{2}\pm\,e^{-ik_{1}a}\cos\frac{\theta_{1}}{2} &  -e^{ik_{2}a}\cos\frac{\theta2}{2}\pm\,e^{-ik_{2}a}\sin\frac{\theta2}{2} & -R\times\, &-e^{-ik_{x}a-k_{y}a}\\
  ~&~&e^{i(k_{x}a+\Phi)-k_{y}a }&\\
 ik_{1}\left(e^{ik_{1}a}\cos\frac{\theta_{1}}{2}\mp\,e^{-ik_{1}a}\sin\frac{\theta_{1}}{2}\right)  & ik_{2}\left(e^{ik_{2}a}\sin\frac{\theta2}{2}\pm\,e^{-ik_{2}a}\cos\frac{\theta2}{2}\right)  &  -(ik_{x}-k_{y})\times &R(ik_{x}+k_{y})\times\\
  ~&~&e^{ik_{x}a-k_{y}a}&e^{-i(k_{x}a+\Phi)-k_{y}a}\\
ik_{1}\left(e^{ik_{1}a}\sin\frac{\theta_{1}}{2}\mp\,e^{-ik_{1}a}\cos\frac{\theta_{1}}{2}\right) &-ik_{2}\left(e^{ik_{2}a}\cos\frac{\theta2}{2}\pm\,e^{-ik_{2}a}\sin\frac{\theta2}{2}\right)&-R(ik_{x}-k_{y})\times&(ik_{x}+k_{y})\times\\
 ~&~&e^{i(k_{x}a+\Phi)-k_{y}a }&e^{-ik_{x}a-k_{y}a}
\end{array}
\right).
\end{equation}
In the energy region $-\Delta\le\,E\le-\frac{\Delta^{2}}{2m\alpha^{2}}$, the matrix $\textbf{M}$ reads
\begin{equation}
M_{\pm}=\left(
\begin{array}{cccc}
 -e^{-\Gamma\,a-i\varphi}\pm\,e^{\Gamma\,a} &  e^{ika}\sin\frac{\theta}{2}\mp\,e^{-ika}\cos\frac{\theta}{2} & -e^{ik_{x}a-k_{y}a}  & -R\times\\
 ~&~&~&e^{-i(k_{x}a+\Phi)-k_{y}a}\\
e^{-\Gamma\,a}\mp\,e^{\Gamma\,a-i\varphi} &  -e^{ika}\cos\frac{\theta}{2}\pm\,e^{-ika}\sin\frac{\theta}{2} & -R\times\, &-e^{-ik_{x}a-k_{y}a}\\
  ~&~&e^{i(k_{x}a+\Phi)-k_{y}a }&\\
 \Gamma\left(e^{-\Gamma\,a-i\varphi}\pm\,e^{\Gamma\,a}\right) & ik\left(e^{ika}\sin\frac{\theta}{2}\pm\,e^{-ika}\cos\frac{\theta}{2}\right)  &  -(ik_{x}-k_{y})\times &R(ik_{x}+k_{y})\times\\
  ~&~&e^{ik_{x}a-k_{y}a}&e^{-i(k_{x}a+\varphi)-k_{y}a}\\
-\Gamma\left(e^{-\Gamma\,a}\pm\,e^{\Gamma\,a-i\varphi}\right) &-ik\left(e^{ika}\cos\frac{\theta}{2}\pm\,e^{-ika}\sin\frac{\theta}{2}\right)&-R(ik_{x}-k_{y})\times&(ik_{x}+k_{y})\times\\
 ~&~&e^{i(k_{x}a+\Phi)-k_{y}a }&e^{-ik_{x}a-k_{y}a}
\end{array}
\right).
\end{equation}
In the energy region $-\frac{\Delta^{2}}{2m\alpha^{2}}\le\,E\le\Delta$, the matrix $\textbf{M}$ reads
\begin{equation}
M_{\pm}=\left(
\begin{array}{cccc}
 e^{-\Gamma\,a+i\varphi}\pm\,e^{\Gamma\,a} &  e^{ika}\sin\frac{\theta}{2}\mp\,e^{-ika}\cos\frac{\theta}{2} & -e^{ik_{x}a-k_{y}a}  & -R\times\\
 ~&~&~&e^{-i(k_{x}a+\Phi)-k_{y}a}\\
e^{-\Gamma\,a}\pm\,e^{\Gamma\,a+i\varphi} &  -e^{ika}\cos\frac{\theta}{2}\pm\,e^{-ika}\sin\frac{\theta}{2} & -R\times\, &-e^{-ik_{x}a-k_{y}a}\\
  ~&~&e^{i(k_{x}a+\Phi)-k_{y}a }&\\
 -\Gamma\left(e^{-\Gamma\,a+i\varphi}\mp\,e^{\Gamma\,a}\right) & ik\left(e^{ika}\sin\frac{\theta}{2}\pm\,e^{-ika}\cos\frac{\theta}{2}\right)  &  -(ik_{x}-k_{y})\times &R(ik_{x}+k_{y})\times\\
  ~&~&e^{ik_{x}a-k_{y}a}&e^{-i(k_{x}a+\varphi)-k_{y}a}\\
-\Gamma\left(e^{-\Gamma\,a}\mp\,e^{\Gamma\,a+i\varphi}\right) &-ik\left(e^{ika}\cos\frac{\theta}{2}\pm\,e^{-ika}\sin\frac{\theta}{2}\right)&-R(ik_{x}-k_{y})\times&(ik_{x}+k_{y})\times\\
 ~&~&e^{i(k_{x}a+\Phi)-k_{y}a }&e^{-ik_{x}a-k_{y}a}
\end{array}
\right).
\end{equation}
Here $k_{x}=k_{\rho}\cos\phi$, $k_{y}=k_{\rho}\sin\phi$, and $M_{\pm}$ means $M_{\sigma^{x}\mathcal{P}=\pm1}$.
\end{widetext}

\end{document}